\documentclass[aps,prb]{revtex4}
\usepackage{amsmath}
\usepackage{graphicx}

\begin{document}

\title{Fluctuation-induced  friction and heat transfer at the water-multilayer graphene interface}

\author{A. I. Volokitin $^{*}$}

\affiliation{Samara State Technical University,  443100 Samara, Russia}

\begin{abstract}
Calculations of friction and heat transfer at the water-multilayer graphene interface using the theories of phononic and radiative friction and heat transfer are presented. The phononic contributions to friction and heat transfer are many  orders of magnitude larger than the radiative contributions. Phononic friction and heat transfer slightly increase with an increase in the number of graphene layers $N$ and reach saturation at $N>5$, which is associated with an increase in the phonon transmission coefficient through the interface and a finite phonon mean free path in the direction perpendicular to the surface. The radiative contributions are almost independent on $N$, since for  distance between water and graphene of the order of the interlayer distance  in graphene the interaction of evanescent  waves with multilayer graphene is limited by the first graphene layer. The results for the phonon contributions agree with the results obtained for the Kapitsa resistance using MD simulation and with the experimental data obtained for the friction coefficients at the water-monolayer graphene interface. The potential difference leads to a strong increase in the radiative contributions to friction and heat transfer, which become approximately an order of magnitude greater than the phononic contributions at a potential difference of $\sim 10$V.
\end{abstract}
\maketitle

PACS: 44.40.+a, 63.20.D-, 78.20.Ci

\section{Introduction}

\vskip 5mm

Friction at the water-carbon interface in a nano size channel has received considerable attention recently due to its relevance in nanoscale systems\cite{Bocquet2020NatMat}.  Experiments and simulations have found that water moves practically without friction through carbon nanotubes\cite{Hummer2001Nature,Majumder2005Nature,Holt2006Science,Whitby2008NanoLett}. These observations have stimulated active studies  in nanotube-based membranes for applications, including
desalination, nano-filtration and energy harvesting\cite{Nair2012Science,Joshi2014Science,Park2014ChemSocRev,Liu2010Science,Siria2013Nature,Geng2014Nature}. However, the mechanism of water-carbon friction remains not entirely clear \cite{Faucher2019JChemPhys,Bocquet2007SoftMat,Thomas2008NanoLet,Falk2010NanoLett,Tocci2014NanoLett}.

Fluctuations inside the media produce a fluctuating electromagnetic field, which is responsible for the Casimir forces. At non-equilibrium conditions, when there is  a temperature difference between the media, or the media are moving relative to each other, the same fluctuating electromagnetic field produces radiative heat transfer and Casimir friction\cite{Volokitin2007RMP,Volokitin2017Book}. In the near field, for the separation between media smaller than the characteristic thermal wavelength $\lambda_T$ ($\lambda_T=c\hbar/k_BT,$ at room temperature $\lambda_T\sim 10\mu$m) radiative heat transfer and Casimir friction are enhanced  by many orders of the magnitude due to the contribution of evanescent electromagnetic waves. However, in an extreme near-field at a separation between media $\sim 1$nm,  van der Waals and electrostatic interaction between fluctuating surface displacements produce phononic heat transfer dominated by acoustic waves, which exceeds radiative heat transfer\cite{Persson2011JPCM,Volokitin2019JETPLett,Volokitin2020JPCMa,Volokitin2020JPCMb}.

Here we present a general theory of phononic friction between closely spaced media,  which is a generalization of our theory of phononic heat transfer\cite{Persson2011JPCM}.  In our theory, friction arises from the interaction between surface displacements that experience  thermal and quantum fluctuations.  The interaction between surfaces was described by the Lennard-Jones potential and the electrostatic potential difference. The theory is applied to study friction between water flow and graphene layers. Also given are the results of calculations using radiative  theories of friction (Casimir friction) and heat transfer mediated by  a fluctuating electromagnetic field.  The importance of the Casimir friction between fluid flow and two-dimensional structures was first demonstrated  in Ref.\cite{Volokitin2008PRB}.  The phononic and radiative friction and heat transfer between two gold surfaces was studied   recently in Refs.\cite{Volokitin2021PRB,Volokitin2021ASSA}.  It was found that a strong enhancement  in radiative friction and heat transfer is expected in the presence of an electric double layer on gold-gold  interface\cite{Volokitin2021PRB}.

\begin{figure}
\includegraphics[width=0.5\textwidth]{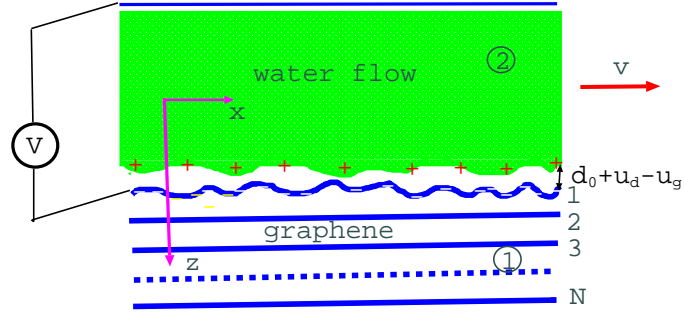}
\caption{Schematic view of water flow  and graphene layers. Thermal fluctuations of surface displacements of water $u_d$ and graphene sheet $u_g$ produce fluctuating stresses acting on the surfaces due to the van der Waals and electrostatic interaction. These fluctuating stresses are responsible for the phononic friction and heat transfer between water flow and graphene layers. 
\label{scheme}}
\end{figure}

\section{Theory}

\subsection{Phononic friction and heat transfer}

A schematic view of our model, which consists of a water block and a system of $N$ graphene layers, is shown in Fig. \ref{scheme}.   Calculations of the Kapitza resistance within a similar model using molecular dynamics methods showed\cite{Alosious2020JChemP,Alexeev2015NanoLett} that the  results depend on the number of graphene layers, but do not depend on the thickness of the water block when it exceeds 2nm. In our calculations, the water block was considered in the isotropic continuum model, and the graphene sheet was considered as an elastic membrane.
 The interlayer carbon interactions were modelled by pairwise Lennard-Jones (L.-J) potential
\begin{equation}
V_{ij}=4\varepsilon_{ij}\left[\left(\frac{\sigma_{ij}}{r_{ij}}\right)^{12}-
\left(\frac{\sigma_{ij}}{r_{ij}}\right)^{6}\right]
\label{via}
\end{equation}
using the parameters from Girifalco \textit{et al.}\cite{Girifalco} The carbon-water interaction were modelled by L.-J potentials with parameters taken from work of Werden\textit{et al.}\cite{Werder} Values of L.-J parameters are provided in Table \ref{parameters}

\begin{table}[ht]
\caption{Lennard-Jones parameters} 
\centering 
\begin{tabular}{c| c c }
\hline\hline 
Pair &  $\sigma$({\AA}) & $\varepsilon$(kJ/mol)  \\ [0.5ex]
\hline 
C-O\cite{Werder}& 3.190 & 0.3920\\
C-C\cite{Girifalco}& 3.414 & 0.2313\\
C-H\cite{Werder} & 0 & 0\\  
\hline 
\end{tabular}
\label{parameters} 
\end{table}

Thermal fluctuations of surface displacements of  water $u_d$ and graphene $u_g$ will create surface stresses. In the rest reference frame  of graphene ($K$ frame), in which water moves at velocity $\mathbf{v}$, a stress  acts on the graphene surface \cite{Volokitin2020JPCMb}
\begin{equation}
\sigma_g(\mathbf{x}, t)=-K_0u_g(\mathbf{x},t)+\int d^2\mathbf{x}_1K(\mathbf{x}-\mathbf{x}_1)u_d(\mathbf{x}_1-\mathbf{v}t, t),
\label{sigmag}
\end{equation}
and in    the co-moving   to   water flow reference  frame ($K^{\prime}$ frame) on  water surface  acts  stress
\begin{equation}
\sigma_d(\mathbf{x}^{\prime},t)=-K_0u_d(\mathbf{x}^{\prime},t)+\int d^2\mathbf{x}_1^{\prime}K(\mathbf{x}^{\prime}-\mathbf{x}^{\prime}_1)u_g(\mathbf{x}^{\prime}_1+\mathbf{v}t, t).
\label{sigmad}
\end{equation}
In the case of L.-J potential 
\[
K_0=30\pi n_On_C\varepsilon_{C-O}d_0,\,\, K((\mathbf{x}-\mathbf{x}_1)=\frac{4K_0}{5\pi}\left\{\frac{2\sigma_{C-O}^{12}}{[(\mathbf{x}-\mathbf{x}_1)^2+
d_0^2]^7}-\frac{\sigma_{C-O}^6}{[(\mathbf{x}-\mathbf{x}_1)^2+
d_0^2]^4}\right\}
\]
where $d_0=(2/5)^{1/6}\sigma_{C-O}=2.738${\AA} is the equilibrium distance between water and graphene, $n_O=3.34\cdot 10^{28}$m$^{-3}$ is the concentration of oxygen atoms in water, $n_C=3.85\cdot 10^{19}$m$^{-2}$ is the concentration of carbon atoms in graphene sheet. As a result of the Fourier transformation 
\begin{equation}
u_i(\mathbf{x},t)=\int\frac{d\omega}{2\pi}\int\frac{d^2\mathbf{x}}{(2\pi)^2}u_i(\omega,q)e^{-i\omega t+i\mathbf{q}\cdot \mathbf{x}},
\end{equation}
Eqs. (\ref{sigmag}) and (\ref{sigmad}) take the form 
\begin{equation}
\sigma_g(\omega,q)=-K_0u_g(\omega,q) + K_1u_d(\omega^{\prime},q),
\end{equation}
\begin{equation}
\sigma_d(\omega^{\prime},q)=-K_0u_d(\omega^{\prime},q) + K_1u_g(\omega,q),
\end{equation}
where $\omega^{\prime}=\omega - q_xv$ is the Doppler shifted frequency,
\begin{equation}
K_1=\frac{q^3d_0^3K_0}{12}\left[\frac{q^3d ^3_0K_6(qd_0)}{192}-K_3(qd_0)\right],
\label{2bvdw}
\end{equation} 
where the value of the following integral was used\cite{Handbook}
\begin{equation}
\int d^2\mathbf{x}\frac{e^{i\mathbf{q}\cdot\mathbf{x}}}{(r^2+
d^2)^{\mu +1}}=2\pi\int_0^{\infty}\frac{J_0(qr)r dr}{(r^2+d^2)^{\mu+1}}=\frac{\pi}{2^{\mu-1}} 
\left(\frac{q}{d}\right)^{\mu}\frac{K_{\mu}(qd)}{\Gamma(\mu +1)}
\end{equation}
where $K_{\mu}(z)$ is the Bessel function of the second kind and order $\mu$ (see Ref.\cite{Handbook}).

A potential difference between graphene and electrode (see Fig.\ref{scheme}) induces the surface charge densities on graphene $\sigma_g=E_0/4\pi$ and water $\sigma_d=-\sigma_g(\varepsilon_{d0}-1)/\varepsilon_{d0}\approx \sigma_g$ where $E_0$ is the electric field in the vacuum gap between graphene and water induced by voltage and $\varepsilon_{d0}\approx 80$ is the static dielectric constant for water. In the presence of a potential difference between graphene and water a  electrostatic interaction between fluctuating displacements of charged surfaces of graphene and water also contributes to the stresses which are determined by \cite{Volokitin2020JPCMb} 
\begin{equation}
\sigma_g=-K_gu_g+Ku_d,
\label{sigmag}
\end{equation}
\begin{equation}
\sigma_d=-K_{d}u_d+Ku_g,
\label{sigmad}
\end{equation}
where 
\begin{equation}
K_{g}=-\frac{E_0^2}{4\pi}\frac{q\left(1+e^{-2qd}R_{d0}\right)}{1-e^{-2qd}R_{d0}}+K_0,
\end{equation}
\begin{equation}
K_{d}=-\frac{E_0^2}{4\pi}\frac{qR_{d0}\left(1+e^{-2qd}\right)}{1-e^{-2qd}R_{d0}}+K_0,
\end{equation}
\begin{equation}
K=-\frac{E_0^2}{2\pi}\frac{qe^{-qd}R_{d0}}{1-e^{-2qd}R_{d0}}+K_1.
\end{equation}
where 
\begin{equation}
R_{d0}=\frac{\varepsilon_{d0}-1}{\varepsilon_{d0}+1}
\end{equation}
where $\varepsilon_{d0}$ is the static dielectric constant for a dielectric.

The   surface displacements  due to thermal and quantum fluctuations are determined by \cite{Persson2011JPCM,Volokitin2019JETPLett,Volokitin2020JPCMa,Volokitin2020JPCMb}
\begin{equation}
u_g(\omega)= u_g^f(\omega)+M_g(\omega)[-K_gu_g(\omega)+Ku_d(\omega^{\prime})],
\label{ug}
\end{equation}
\begin{equation}
u_d(\omega^{\prime})= u_d^f(\omega^{\prime})+M_d(\omega^{\prime})[-K_du_d(\omega^{\prime})+Ku_g(\omega)],
\label{ud}
\end{equation}
where according to the fluctuation-dissipation theorem, the spectral density of fluctuations of the surface displacements is determined by  \cite{LandauStatisticalPhysics}
\begin{equation}
\langle|u_i^f|^2\rangle = \hbar \mathrm{Im}M_i(\omega,q)\coth\frac{\hbar\omega}{2k_BT_i}
\label{fdt}
\end{equation}
where $<\cdots >$ denotes thermal average, $M_i$ is the mechanical susceptibility for surface $i$: $u_i=M_i\sigma$.

From Eqs. (\ref{ug}) and  (\ref{ud})
\begin{equation}
u_g(\omega) = \frac{(1+K_0M_d(\omega^{\prime})u_g^f(\omega)+KM_g(\omega)u_d(\omega^{\prime})}{(1+K_gM_g(\omega))(1+K_dM_d(\omega^{\prime}))-K^2M_g(\omega)M_d(\omega^{\prime})},
\label{ugg}
\end{equation}
\begin{equation}
u_d(\omega^{\prime}) = \frac{(1-K_gM_g(\omega)u_d^f(\omega^{\prime})+KM_d(\omega^{\prime}))u_g(\omega)}{(1+K_gM_g(\omega))(1+K_dM_d(\omega^{\prime}))-K^2M_g(\omega)M_d(\omega^{\prime})},
\label{udd}
\end{equation}
The heat generated by fluctuating stresses in water and graphene  are determined by 
\[
\left(\begin{array}{c}
\dot{Q}_g\\
\dot{Q_d}
\end{array}\right) = \int\frac{d\omega}{2\pi}\int\frac{d^2\mathbf{q}}{(2\pi)^2}\left(\begin{array}{c}
-i\omega<u_g(\omega)\sigma_g(\omega)>\\
-i\omega^{\prime}<u_d(\omega^{\prime})\sigma_d(\omega^{\prime})>
\end{array}\right)=
\]
\begin{equation}
= 2\hbar\int\frac{d\omega}{2\pi}\int\frac{d^2\mathbf{q}}{(2\pi)^2}\left(\begin{array}{c}
\omega\\
-\omega^{\prime}
\end{array}\right)
\frac{K^2\mathrm{Im}M_g(\omega)M_d(\omega^{\prime})[n_d(\omega^{\prime})-n_g(\omega)]}{\left|(1+K_gM_g(\omega))(1+K_dM_d(\omega^{\prime}))-K^2M_g(\omega)M_d(\omega^{\prime})\right|^2}
\label{heat}
\end{equation}
where $n_i(\omega)=[\mathrm{exp}(\hbar\omega/k_BT_i)-1]^{-1}$. The friction force can be calculated from equation\cite{Volokitin2017Book}
\begin{equation}
f_xv=\dot{Q}_g + \dot{Q}_d.
\label{fv}
\end{equation}
From Eqs.(\ref{heat}-\ref{fv}) the friction force\cite{Volokitin2021ASSA} 
\begin{equation}
f_x=4\hbar\int_{0}^\infty\frac{d\omega}{2\pi}\int\frac{d^2q}{(2\pi)^2}q_x
\frac{K^2\mathrm{Im}M_g(\omega)\mathrm{Im}M_d(\omega^{\prime}) }{\mid (1+K_gM_g(\omega) )(1+K_dM_w(\omega^{\prime} )-K^2M_g(\omega)M_d(\omega^{\prime})\mid^2}\left[n_d(\omega^{\prime})-n_g(\omega)\right].
\label{frforce}
\end{equation}
To linear order on velocity $v$ the friction force $f_x=\gamma v$ where at $T_g=T_d=T$ the phonon friction coefficient  
\begin{equation} 
\gamma_{ph}=\frac{\hbar^2}{8\pi^2k_BT} \int_0^\infty \frac{d\omega}{\mathrm{sinh}^2(\hbar\omega/2k_BT)}\int_{0}^{\infty}dqq^3\frac{K^2\mathrm{Im}M_g(\omega)\mathrm{Im}M_d(\omega) }{\mid (1+K_gM_g(\omega) )(1+K_dM_d(\omega)-K^2M_g(\omega)M_d(\omega)\mid^2}.
\label{frcoef}
\end{equation}
At $v=0$ the heat flux due to the phonon tunneling
\begin{equation}
J(T_g, T_d)=\dot{Q_g}=-\dot{Q_d}= 4\hbar\int_0^{\infty}\frac{d\omega}{2\pi}\int\frac{d^2\mathbf{q}}{(2\pi)^2}
\omega
\frac{K^2\mathrm{Im}M_g(\omega)M_d(\omega)[n_d(\omega)-n_g(\omega)]}{\left|(1+K_gM_g(\omega))(1+K_dM_d(\omega))-K^2M_g(\omega)M_d(\omega)\right|^2}
\label{heat}
\end{equation}
and the heat transfer coefficient 
\[
\alpha_{ph}(T)=\lim_{\Delta T\rightarrow 0}\frac{J(T, T+\Delta T)}{\Delta T}
\]
\begin{equation}=\frac{\hbar^2}{4\pi^2k_BT^2} \int_0^\infty \frac{d\omega\omega^2}{\mathrm{sinh}^2(\hbar\omega/2k_BT)}\int_{0}^{\infty}dqq\frac{K^2\mathrm{Im}M_g(\omega,q)\mathrm{Im}M_d(\omega,q) }{\mid (1+K_gM_g(\omega,q) )(1+K_dM_d(\omega,q)-K^2M_g(\omega,q)M_d(\omega,q)\mid^2}.
\label{conduct}
\end{equation}

\subsection{Casimir friction and radiative heat transfer}

According to the theories of the Casimir friction  and radiative heat transfer\cite{Volokitin2011PRL,Volokitin2011PRB,Volokitin2020JPCMb} the Casimir friction and heat transfer  coefficients are determined by
\begin{equation} 
\gamma_{rad}=\frac{\hbar^2}{8\pi^2k_BT} \int_0^\infty \frac{d\omega}{\mathrm{sinh}^2(\hbar\omega/2k_BT)}\int_{0}^{\infty}dqq^3\frac{\mathrm{Im}R_g(\omega)\mathrm{Im}R_d(\omega)e^{-2qd} }{\mid 1-e^{-2qd}R_g(\omega)R_d(\omega)\mid^2},
\label{frcoefrad}
\end{equation}
\begin{equation}
\alpha_{rad}(T)=\frac{\hbar^2}{4\pi^2k_BT^2} \int_0^\infty \frac{d\omega\omega^2}{\mathrm{sinh}^2(\hbar\omega/2k_BT)}\int_{0}^{\infty}dqq\frac{\mathrm{Im}R_g(\omega,q)\mathrm{Im}R_d(\omega,q) e^{-2qd}}{\mid {\mid 1-e^{-2qd}R_g(\omega)R_d(\omega)\mid^2}\mid^2},
\label{heatcoefrad}
\end{equation}
where $R_g$ and $R_d$ are the reflection amplitudes for the graphene and dielectric surfaces.

\section{Numerical results}

In the  elastic continuum model for isotropic medium  the surface displacement under the action of external mechanical stress $u=M\sigma_{zz}^{ext}$ where mechanical susceptibility
\cite{Persson2001JPCM}
\begin{equation}
M_d=\frac{i}{\rho c_t^2}\left(\frac{\omega}{c_t}\right)^2\frac{p_l(q,\omega)}{S(q,\omega)},
\label{mecsucs}
\end{equation}
where
\[
S(q,\omega)=\left[\left(\frac{\omega}{c_t}\right)^2-2q^2\right]^2+4q^2p_tp_l,
\]
\[
p_t=\left[\left(\frac{\omega}{c_t}\right)^2-q^2+i0\right]^{1/2}, \,\,p_l=\left[\left(\frac{\omega}{c_l}\right)^2-q^2+i0\right]^{1/2},
\]
where  $\rho$,  $c_l$, and
$c_t$ are the mass density of the medium, the velocity of the longitudinal and transverse acoustic waves. For water $c_t=0$, $c_l=1500$m/s and $M_d$ is reduced to the form
\begin{equation}
M_d=\frac{ip_l}{\rho \omega^2}.
\label{mecsucsl}
\end{equation}

One graphene layer can be considered as 
an elastic membrane  for which the mechanical susceptibility related with out-of-plane displacement\cite{Persson2011JPCM}
\begin{equation}
M_{1g}=\frac{1}{\kappa q^4-\rho \omega^2-i\omega \rho\gamma},
\end{equation}
where the bending stiffness of graphene    $\kappa\approx 1$eV, $\rho=7.7\cdot 10^{-7}$kg/m$^2$ is the surface mass density of   graphene, $\gamma$ is the damping constant for flexural motion of graphene which was estimated in Ref.\cite{KapitzRes2016PRB} as
\begin{equation}
\gamma=\frac{\omega T}{\alpha T_{RT}}
\end{equation}
where $T_{RT}=300$K is the room temperature. For number of interacting graphene layers $N\ge 2$ the mechanical susceptibility is calculated in Appendix and is given by 
\begin{equation}
M_{Ng}=-\frac{\Delta(1-\lambda^{2(N-1)})-K_c(1-\lambda)(1+\lambda^{2N-3})}
{\Delta^2(1-\lambda^{2(N-1)})-2\Delta K_c(1-\lambda)(1+\lambda^{2N-3})+
K_c^2(1-\lambda)^2(1-\lambda^{2(N-2)})}
\end{equation}
where $\Delta =  \rho\omega^2-\kappa q^4 +i\omega\eta(\omega)$, for the L.-J. interaction between graphene layers the spring constant for interlayer interaction 
\begin{equation}
K_c=8\pi n_c^2\varepsilon_{c-c}\left[11\left(\frac{\sigma_{c-c}}{a}\right)^{12}-5\left(\frac{\sigma_{c-c}}{a}\right)^6\right]=1.16\cdot 10^{20} \mbox{Nm}^{-3}
\end{equation}
where $a=3.35${\AA} is the interlayer separation, $n_c=3.85\cdot10^{19}$m$^{-3}$ is the concentration of carbon atom in graphene sheet,
\begin{equation}
\lambda=1-\frac{\Delta}{2K_c}+\sqrt{\left(\frac{\Delta}{2K_c}-1\right)^2-1}
\end{equation}
For $|\Delta/K_c|\ll 1$ $\gamma\rightarrow 1$,   the continuous medium approximation can be used for which 
\begin{equation}
M_{gl}=-\frac{\mathrm{cot}\,pl}{\sqrt{\Delta K_c}}
\end{equation}
where $p=\sqrt{\Delta/K_c}/a$,  $l=Na$ is the thickness of he graphene block.

The calculations of the reflection amplitude for multilayer graphene for arbitrary number of graphene layers $N$ are given in Appendix \ref{B}. For monolayer graphene in the presence of the potential difference between graphene and water\cite{Volokitin2019JETPLett}
\begin{equation}
R_{g1}=\frac{\varepsilon_g -1 +2\pi q\sigma_g^2M_g}{\varepsilon_g(1  -2\pi q\sigma_g^2M_g)},
\label{rcg1}
\end{equation}
and for dielectric 
\begin{equation}
R_d=\frac{\varepsilon_d -1 +4\pi q\sigma_d^{ 2}M_d\varepsilon_d}{\varepsilon_d+1  -4\pi q\sigma_d^{2}M_d\varepsilon_d}
\label{rcd1}
\end{equation}
where $\varepsilon_g$ and $\varepsilon_d$ are the dielectric function of graphene and water.
For multilayer graphene ($N>1$) according to Eq.(\ref{RgN})
\begin{equation}
R_{gN}=\frac{e^{2qa}R_1R_2(1-\lambda^{2N})}{R_2-R_1\lambda^{2N}}
\end{equation}
where 
\begin{equation}
R_1=\frac{ e^{-qa}-\lambda}{ e^{qa}-\lambda},\,\,R_2=\frac{ \lambda e^{-qa}-1}{ \lambda e^{qa}-1},
\end{equation}
\begin{equation}
\lambda=e^{-qa}+\varepsilon_g\frac{e^{qa}-e^{-qa}}{2}-\sqrt{\left[e^{-qa}+\varepsilon_g\frac{e^{qa}-e^{-qa}}{2}\right]^2-1}.
\label{root}
\end{equation}

In the numerical calculations we used the dielectric function of graphene, which was
calculated  within the random-phase approximation (RPA)
\cite{Wunsch2006NJP,Hwang2007PRB}.  
\begin{equation}
\varepsilon_g(\omega,q)=1+\frac{8k_Fe^2}{\hbar
v_Fq}-\frac{e^2q}{\hbar \sqrt{\omega^2-v_F^2q^2}}\Bigg \{G\Bigg
(\frac{\omega+2v_Fk_F}{v_Fq}\Bigg )- G\Bigg
(\frac{\omega-2v_Fk_F}{v_Fq}\Bigg )-i\pi \Bigg \},
\end{equation}
where
\begin{equation}
G(x)=x\sqrt{x^2-1} - \ln(x+\sqrt{x^2-1}),
\end{equation}
where the Fermi wave vector $k_F=(\pi n_g)^{1/2}$, $n_g$ is the
concentration of charge carriers, the Fermi energy
$\epsilon_F=\hbar v_Fk_F$,  $v_F\approx 10^6$ m/s is the Fermi velocity.

From    the recurrence relation (\ref{recrel})
\begin{equation}
R_{gN}=\frac{\varepsilon_g-1+e^{-2qa}R_{g(N-1)}(2-\varepsilon_g)}{\varepsilon_g-(\varepsilon_g-1)e^{-2qa}R_{g(N-1)}}.
\label{recrel1}
\end{equation}
follows that $R_{gN}$ is reduced to $R_1$ for $\mathrm{exp}(-2qa)\ll 1$ and when $R_{g1}$ has no singularities. In Eqs.(\ref{frcoefrad}) and (\ref{heatcoefrad}) the main contribution give $q\sim 1/d$ thus for small separation, when $d\approx a$, the first condition is fulfilled and the graphene dielectric function has no singularities at $q\sim 1/d\gg k_F$. Thus for water-multilayer graphene interface the reflection amplitude is determined approximately by $R_{g1}$ and the dependence of the Casimir friction and heat transfer coefficients on the number of graphene layers is negligible. This results is related with quick decay of the evanescent waves with increasing distance from the interface and was confirmed by numerical calculations using the reflection amplitudes $R_{gN}$ from Eq.(\ref{RgN}). 

Water has
an extremely large static dielectric function of around 80. The
low frequency contribution to the dielectric function, responsible
for this large static value, is due to relaxation of the permanent
dipoles of the water molecules. It can be accurately  described by the
Debye \cite{Debye} theory of rotational relaxation. The theoretical fit of the experimental data is well described by the Debye formula \cite{Kivchar2015SciRep}:
\begin{equation}
\varepsilon(\omega) = \varepsilon_{\infty}+\frac {\varepsilon_0- \varepsilon_{\infty}}{1-i\omega/\omega_0}
\end{equation}
where at $T=300$K $\varepsilon_{\infty}=6.04$, $\varepsilon_0=77.66$  $\omega_0=1.3\cdot 10^{11}$s$^{-1}$. We note that
water has large absorption in the
radio-frequency range at $\omega \sim \omega_0$, and shows in this
region of the spectrum anomalous dispersion.

\begin{figure}
\includegraphics[width=0.5\textwidth]{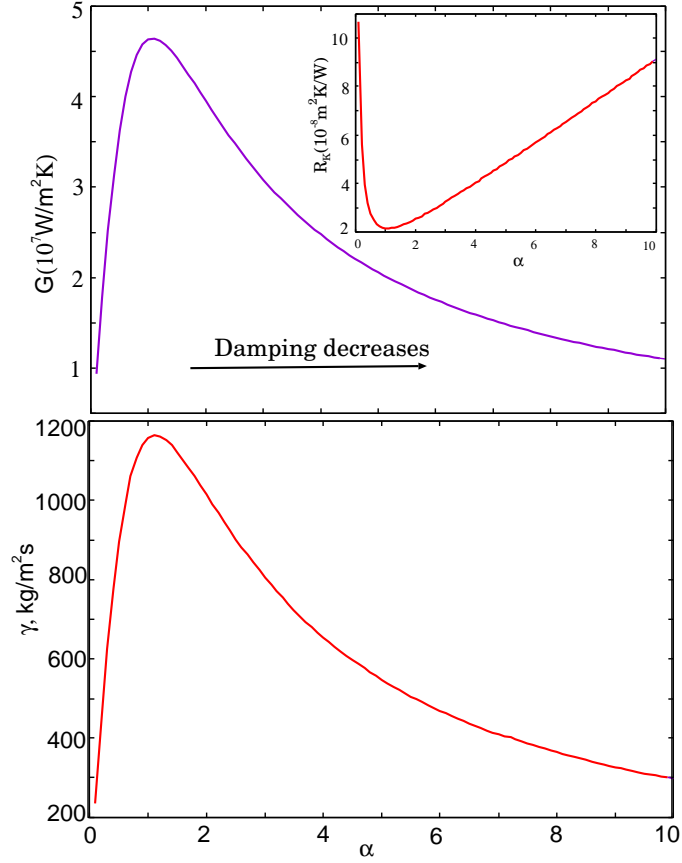}
\caption{Dependence of the phononic heat transfer coefficient $G$ (top) and friction coefficient $\gamma$ (bottom) mediated by   the van der Waals interaction   on the damping parameter $\alpha$ for water-monolayer graphene interface at $T=300$K. Insert in the figure on top shows the Kapitza resistance $R_K=1/G$. An increase of $\alpha$ corresponds to lower damping of the flexural phonons in graphene. 
 \label{vdw_damping}}
\end{figure}

\begin{figure}
\includegraphics[width=0.5\textwidth]{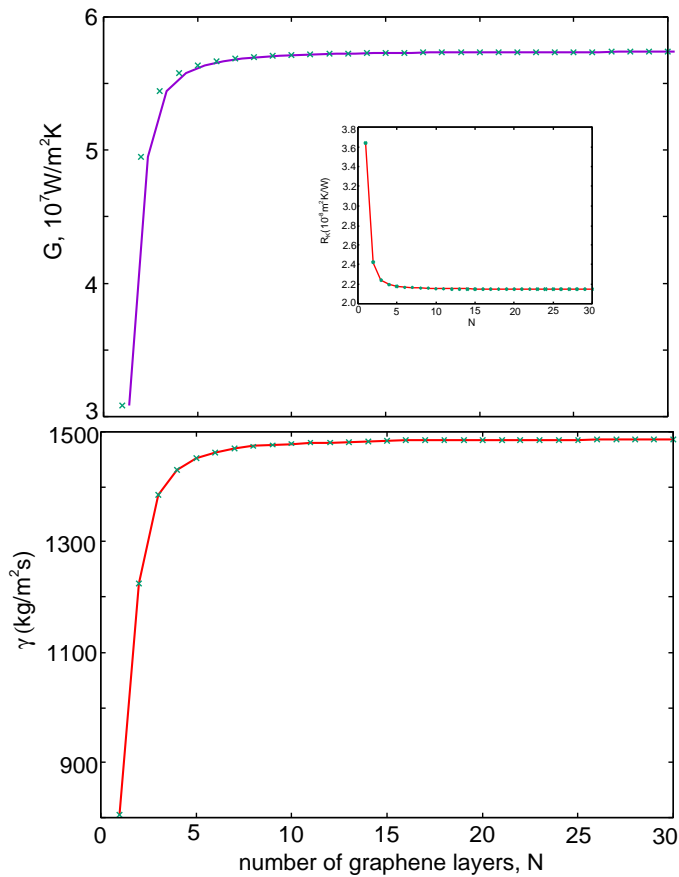}
\caption{Dependence of the phononic heat transfer coefficient $G$ (top) and friction coefficient $\gamma$ (bottom) mediated by   the van der Waals interaction  for water-multilayer graphene interface on the number of graphene layers  at $T=300$K and $\alpha=3$. Insert in the figure on top shows the Kapitza resistance $R_K=1/G$.  \label{vdw_fr_heat_damp}}
\end{figure}

The results of numerical calculations of the dependence of the phononic heat transfer coefficient $G$, the Kapitza resistance $R_K=1/G$)  and friction coefficient $\gamma$  mediated by   the van der Waals interaction   on the damping parameter $\alpha$ for water-monolayer graphene interface at $T=300$K are shown in Fig.\ref{vdw_damping}. The maximum values of $G_{max}=4.5\cdot 10^7$W/m$^2$K, corresponding to the minimum value of  $R_K^{mim}=2\cdot 10^{-8}$m$^2$K/W, and $\gamma_{max}=1200$kg/m$^2$s agree well with results of MD simulations\cite{Alosious2020JChemP,Alexeev2015NanoLett} and experimental data\cite{Secchi2016Nature} for monolayer graphene. Much higher phononic friction coefficient $\sim 10^5$kg/m$^{2}$s for the water-monolayer graphene interface  was calculated using MD simulation in Ref.\cite{Tocci2014NanoLett}. This friction is two orders of magnitudes larger than in our calculations and in experiment\cite{Secchi2016Nature}.  Thus most likely MD simulation overestimate the water-graphene friction coefficient, which is typical in simulation of other water-solid system\cite{Bocquet2010ChemSocRev}.
 
Fig. \ref{vdw_fr_heat_damp} shows dependences of $G$, $R_K$ and $\gamma$ mediated by   the van der Waals interaction   on the number of graphene layers $N$. The $G$ ($R_K$)   increases (decreases) slightly with $N$ and reach saturation for $N>5$ what agrees with the results of MD simulations\cite{Alosious2020JChemP,Alexeev2015NanoLett}. The $\gamma$ also increases  slightly with $N$ what agrees with experimental data\cite{Secchi2016Nature} according to which the friction of water in carbon nanotubes decreases when the radius of nanotube is decreasing. However, in experiment much higher friction ($\gamma \sim 10^4-10^5$kg/m$^2$s)  was observed for graphite. According to the Ref.\cite{Bocquet2021ArXiv}, this increase in friction was attributed to the contribution from surface plasmons that can arise in graphite but this contribution does not exist for multilayer graphene.

\begin{figure}
\includegraphics[width=0.5\textwidth]{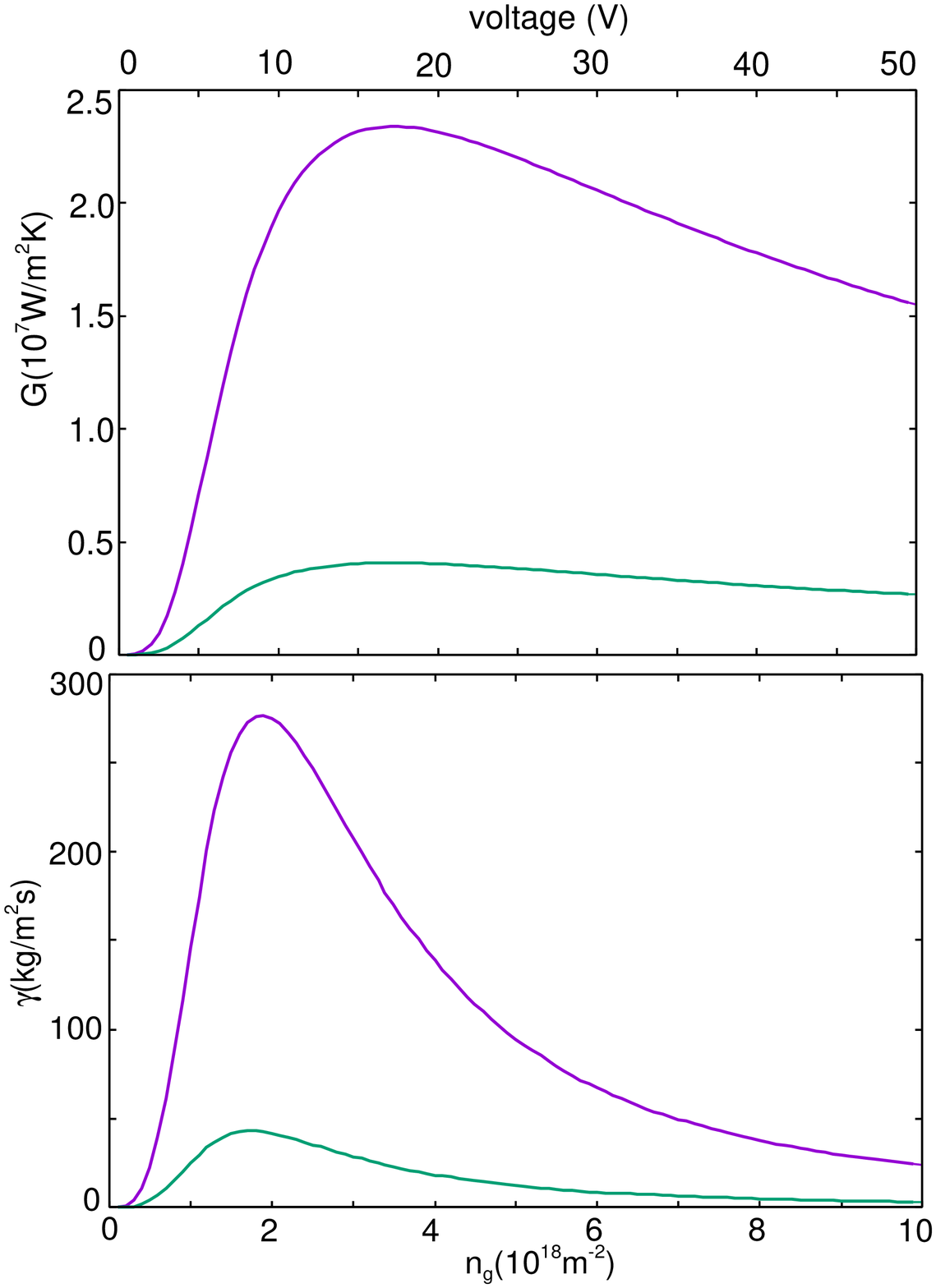}
\caption{Dependence of the phononic heat transfer coefficient $G$ (top) and friction coefficient $\gamma$ (bottom) mediated by   the electrostatic interaction  for water-monolayer graphene interface on the graphene electron concentration $n_g$  at $T=300$K,  $\alpha=1$  and 10 for blue and green lines, respectively.  \label{El_fr_heat_dens}}
\end{figure}

A potential difference $\varphi$ between water and graphene surfaces produces the surface charge density on graphene $\sigma_g=n_ge=\varphi/4\pi d$ and water $\sigma_{d}=-\sigma_g(\varepsilon_{d0}-1)/\varepsilon_{d0}$ where due to large value of the static dielectric constant $\varepsilon_{d0}$ for water $\sigma_g\approx -\sigma_{d}$. The electrostatic interaction between charged surfaces produces additional contributions to $G$ and $\gamma$ which are shown in Fig. \ref{El_fr_heat_dens} in the dependence on the graphene electron concentration $n_g$ and voltage for water-monolayer graphene interface. The maximum values for $G$ and $\gamma$ are obtained with a potential difference $\sim 10$V and they turn out to be several times smaller than the corresponding results for the van der Waals interaction.

\begin{figure}
\includegraphics[width=0.5\textwidth]{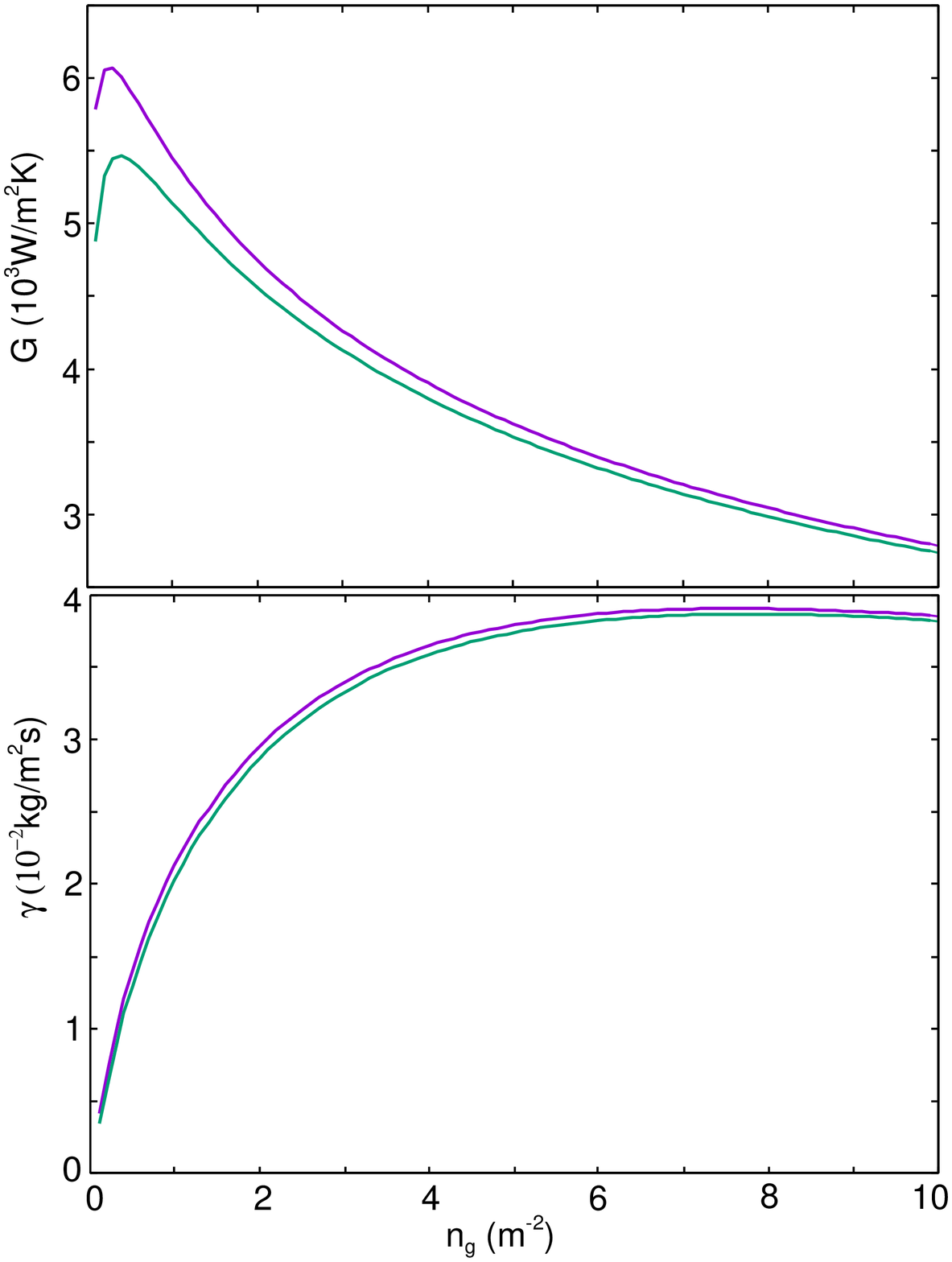}
\caption{Dependence of the radiative heat transfer coefficient $G$ (top) and friction coefficient $\gamma$ (bottom) for water-monolayer graphene interface mediated by   a fluctuating electromagnetic field   for neutral  graphene sheet on the graphene electron concentration  $n_g$  at $T=300$K. Blue and green lines are for $N=\infty $ and $N=1 $, respectively. \label{Rad_fr_heat_concent_sigm=0}}
\end{figure}

Fig. \ref{Rad_fr_heat_concent_sigm=0} shows the dependence of the radiative heat transfer coefficient $G$ (top) and friction coefficient $\gamma$ (bottom) for water-monolayer graphene interface mediated by   a fluctuating electromagnetic field on the concentration of electrons in a graphene sheet  when the concentration of electrons changes as a result of doping while this graphene sheet remains neutral. In this case, the fluctuating electromagnetic field is determined by fluctuations in the bulk polarizability of water and the current density in the graphene sheet, while there are no contributions from acoustic waves in water and bending vibrations of the graphene sheet.The  $\gamma $ and $G$ in this case are  many orders  of magnitude smaller  than for   phononic mechanism. 
\begin{figure}
\includegraphics[width=0.5\textwidth]{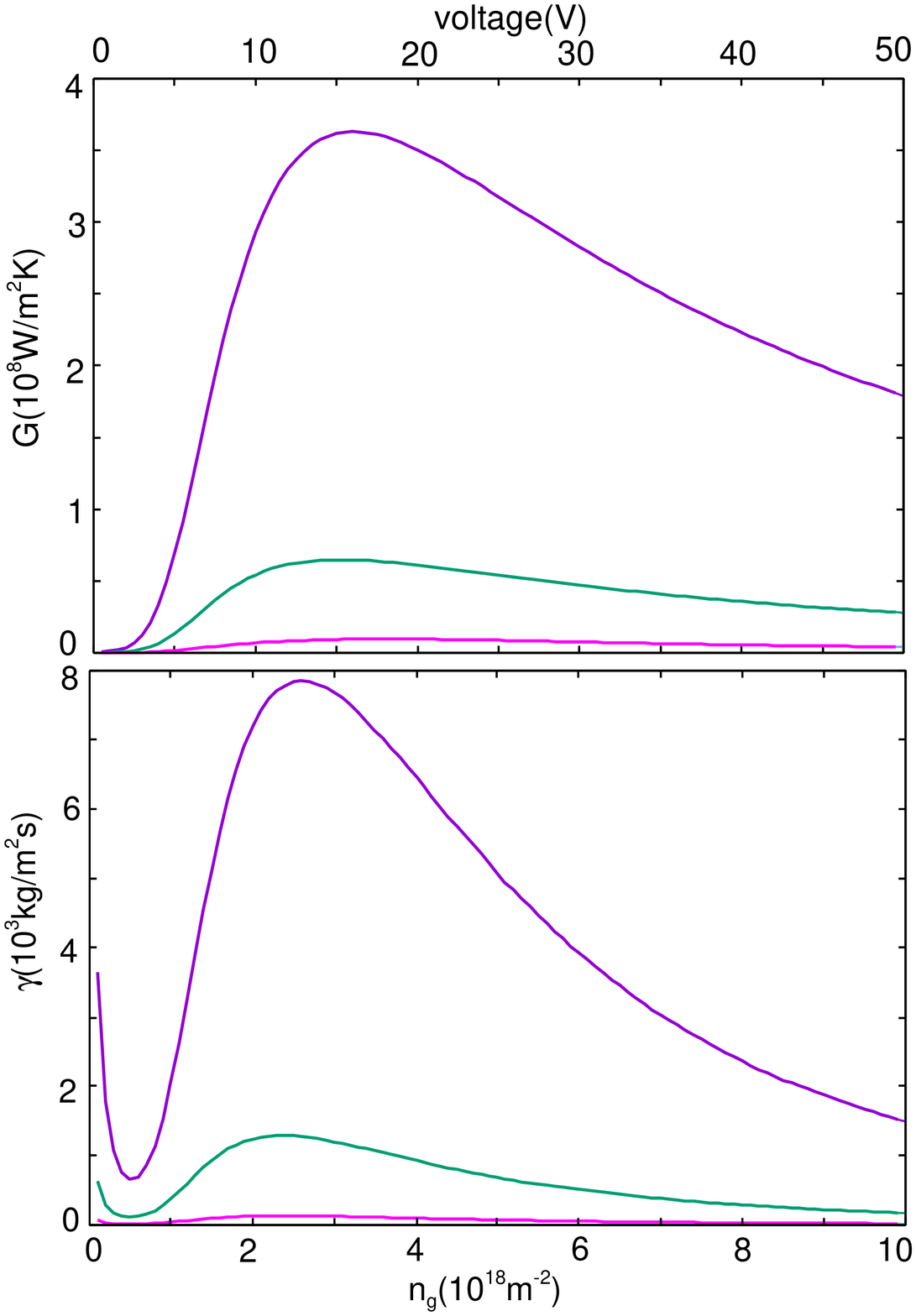}
\caption{Dependence of the radiative heat transfer coefficient $G$ (top) and friction coefficient $\gamma$ (bottom) mediated by   a fluctuating electromagnetic field   for charged water-monolayer graphene interface on the graphene electron concentration $n_g$ which is induced by the potential difference  at $T=300$K. Blue, green and pink line are for the damping parameter for the bending vibrations of the graphene sheet $\alpha=1$, 10 and 100.  \label{Rad_fr_heat_conc}}
\end{figure}

Fig. \ref{Rad_fr_heat_conc}  shows the dependence of the radiative heat transfer coefficient $G$ (top) and friction coefficient $\gamma$ (bottom) for water-monolayer graphene interface mediated by   a fluctuating electromagnetic field on the concentration of electrons in a graphene sheet  when the concentration of electrons is determined  the potential difference between water and graphene which   induces the charge density on the surfaces of water and graphene.  Fluctuations of the  displacements of charged surfaces    create  additional fluctuating electric field, which leads to an increase in friction (heat transfer)  by about an order of magnitude compared to the phonon mechanism at the potential difference $\sim 10$V.

\section{Conclusion}
We have studied friction and heat transfer at the water-multilayer graphene interface interface using  phononic and radiative theories. Unlike MD simulations, our theory does not require laborious numerical calculations. It is shown that the phononic  contributions to friction and heat transfer are many  orders of magnitude larger than the radiative contributions. The radiative contribution rapidly increases with an increase in the potential difference between the surfaces of water and graphene, and it becomes larger than the phonon contribution at a potential difference of the order of 10V. The phonon contributions increase with an increase in the number of graphene layers and reach saturation for N > 5, which is related to the finite phonon mean free path. Radiative contributions are practically independent of N, which is explained by the decay  of evanescent  waves on the interlayer distance $a$ in the case when the separation between water and graphene surfaces $d\approx a$.
Our results   for the phononic contributions are consistent with the results obtained for the Kapitza resistance using MD simulations\cite{Alosious2020JChemP,Alexeev2015NanoLett}  and with  experimental data\cite{Secchi2016Nature} obtained for friction coefficients for  water-monolayer graphene interface, but two orders of magnitude less than the values obtained for friction coefficient using MD simulations\cite{Tocci2014NanoLett}.  Increase in  phononic  $\gamma$  with $N$ agrees with experimental data\cite{Secchi2016Nature} according to which the friction of water in carbon nanotubes decreases when the radius of nanotube is decreasing. However, in experiment much higher friction ($\gamma \sim 10^4-10^5$kg/m$^2$s)  was observed for graphite. According to the Ref.\cite{Bocquet2021ArXiv}, this increase in friction was attributed to the contribution from surface plasmons that can arise in graphite but this contribution does not exist for multilayer graphene. 

\appendix

\section{Mechanical susceptibility of multilayer graphene}

One graphene layer can be considered as 
an elastic membrane  for which the mechanical susceptibility related with out-of-plane displacement\cite{Persson2011JPCM}
\begin{equation}
M_{1g}=\frac{1}{\kappa q^4-\rho \omega^2-i\omega \rho\gamma},
\end{equation}
where the bending stiffness of graphene    $\kappa\approx 1$eV, $\rho=7.7\cdot 10^{-7}$kg/m$^2$ is the surface mass density of   graphene, $\gamma$ is the damping constant for flexural motion of graphene which was estimated in Ref.\cite{KapitzRes2016PRB} as
\begin{equation}
\gamma=\frac{\omega T}{\alpha T_{RT}}
\end{equation}
where $T_{RT}=300$K is the room temperature. For number of interacting graphene layers $N\ge 2$ the mechanical susceptibility can be found from equations

\[
(\Delta - K_c)u_1 + K_cu_2 =-\sigma_0e^{-i\omega + i\mathbf{q}\cdot\mathbf{x}}
\]
\[
K_Cu_1 + (\Delta -2K_c)u_2+K_cu_3 = 0
\]
\[
\cdots\cdots\cdots\cdots\cdots\cdots\cdots\cdots\cdots\cdots\cdots\cdots\cdots
\]
\[
K_cu_{N-2} + (\Delta -2K_c)u_{N-1} +K_cu_N = 0
\]
\begin{equation}
K_cu_{N-1} + (\Delta - K_c)u_N=0
\label{syseq}
\end{equation}
where $\Delta =  \rho\omega^2-\kappa q^4 +i\omega\eta(\omega)$, for the L.-J. interaction between graphene layers the spring constant for interlayer interaction 
\begin{equation}
K_c=8\pi n_c^2\varepsilon_{c-c}\left[11\left(\frac{\sigma_{c-c}}{a}\right)^{12}-5\left(\frac{\sigma_{c-c}}{a}\right)^6\right]=1.16\cdot 10^{20} \mbox{Nm}^{-3}
\end{equation}
where $a=3.35${\AA} is the interlayer separation, $n_c=3.85\cdot10^{19}$m$^{-3}$ is the concentration of carbon atom in graphene sheet. 
The solution of the system of Eqs.(\ref{syseq}) can be written in the form
\begin{equation}
u_n=C_1\lambda^{n-1}+C_2\lambda^{-n+1}
\label{un}
\end{equation}
where  
\begin{equation}
\lambda=1-\frac{\Delta}{2K_c}+\sqrt{\left(\frac{\Delta}{2K_c}-1\right)^2-1}
\end{equation}
is the root of equation 
\begin{equation}
\gamma^2 -2\left(1-\frac{\Delta}{2K_c}\right)\gamma+1=0
\end{equation}
for which $|\gamma|<1$. The constants $C_1$ and $C_2$ are determined by equations which are obtained after substitution of Eq.(\ref{un}) in system (\ref{syseq})
\begin{equation}
[\Delta+K_c(\lambda-1)]C_1+[\Delta+K_c(\frac{1}{\lambda}-1)]C_2=-\sigma_0e^{-i\omega + i\mathbf{q}\cdot\mathbf{x}}
\label{c1}
\end{equation}
\begin{equation}
\lambda^{2(N-1)}\left[\Delta+K_c \left(\frac{1}{\lambda}-1\right)\right]C_1+
\left[\Delta+K_c(\lambda-1)\right]C_2=0
\label{c2}
\end{equation}
From Eqs.(\ref{un})- (\ref{c2})
\begin{equation}
u_1=M_{Ng}\sigma_0e^{-i\omega + i\mathbf{q}\cdot\mathbf{x}}
\end{equation}
where
\begin{equation}
M_{Ng}=-\frac{\Delta(1-\lambda^{2(N-1)})-K_c(1-\lambda)(1+\lambda^{2N-3})}
{\Delta^2(1-\lambda^{2(N-1)})-2\Delta K_c(1-\lambda)(1+\lambda^{2N-3})+
K_c^2(1-\lambda)^2(1-\lambda^{2(N-2)})}
\label{A}
\end{equation}
For $N\rightarrow\infty$
\begin{equation}
\lim_{N\rightarrow\infty}M_{Ng}=-\frac{2}{\Delta + \sqrt{\Delta(\Delta-4K_c)}}
\end{equation}
For $|\Delta/K_c|\ll 1$ $\gamma\rightarrow 1$ thus the continuous medium approximation can be used for which the system (\ref{syseq}) is reduced to equation
\begin{equation}
\Delta u+a^2K_c\frac{d^2u}{d^2z}=0
\label{contap}
\end{equation}
with the boundary conditions
\[
\frac{du}{dz}\Big|_{z=l}=0,\,\,\, aK_c\frac{du}{dz}\Big|_{z=0}=-\sigma_0
\]
where $l=Na$ is the thickness of the system of graphene layers. The solution of Eq.(\ref{contap}) has the form 
\begin{equation}
u(z)=-\frac{\mathrm{cos}\, p(z-l)}{\sqrt{\Delta C}\mathrm{sin}\, pl}\sigma_0
\label{uz}
\end{equation}
where $p=\sqrt{\Delta/K_c}/a$. From (\ref{uz})
\begin{equation}
M_{gl}=u(0)/\sigma_0=-\frac{\mathrm{cot}\,pl}{\sqrt{\Delta K_c}}
\end{equation}
For $l\rightarrow \infty$
\begin{equation}
\lim_{l\rightarrow \infty}M_{gl}=\frac{i}{\sqrt{\Delta K_c}}
\end{equation}

\section{Reflection amplitude of multilayer graphene \label{B}}

In the non-retarded limit, the potential of the electric field of an electromagnetic wave incident on multilayer graphene can be written in the form

\begin{equation}
\varphi(\mathbf{q}), \omega, \mathbf{x}, z) =e^{i\mathbf{q}\cdot\mathbf{x}-i\omega t}\times \left\{
\begin{array}{rl}
e^{-q(z-z_1)}-R_{gN}e^{q(z-z_1)},&  z<0 \\
\cdots\cdots\cdots\cdots\cdots\cdots\cdots,&\\
v_ne^{-q(z_-z_n)}-w_ne^{-q(z-z_n)},& z_n<z<z_{n+1} \\
\cdots\cdots\cdots\cdots\cdots\cdots\cdots\cdots\cdots\\
Te^{-q(z_-z_N)},& z>z_N 
\end{array}\right.
\label{wave}
\end{equation}
The boundary conditions on the surfaces of graphene sheets have the form
\begin{equation}
\begin{array}{rl}
1-R_{gN}=v_1-w_1,\,1+R_{gN}-v_1-w_1=2(\varepsilon_g-1)(v_1-w_1),&   z=z_1 \\
\cdots\cdots\cdots\cdots\cdots\cdots\cdots\cdots\cdots\cdots\cdots\cdots\cdots\cdots\cdots\cdots\\
v_ne^{-qa}-w_ne^{qa}=v_{n+1}-w_{n+1},\,v_ne^{-qa}+w_ne^{qa}-v_{n+1}-w_{n+1}=2(\varepsilon_g-1)(v_{n+1}-w_{n+1}),&   z=z_n \\
\cdots\cdots\cdots\cdots\cdots\cdots\cdots\cdots\cdots\cdots\cdots\cdots\cdots\cdots\cdots\cdots\\
v_{N-1}e^{-qa}-w_{N-1}e^{qa}=T,\,v_{N-1}e^{-qa}+w_{N-1}e^{qa}-T=2(\varepsilon_g-1)T,&   z=z_N 
\end{array}
\label{Radbouncond}
\end{equation}
From Eqs.(\ref{Radbouncond}) for $N=1$
\begin{equation}
R_{g1}=\frac{\varepsilon_g-1}{\varepsilon_g}. 
\label{Rg1}
\end{equation}
Taking into account that for $N>1$
\begin{equation}
w_1=e^{-2qa}R_{g(N-1)}v_1
\end{equation}
from Eqs.(\ref{Radbouncond}) follows the recurrence relation
\begin{equation}
R_{gN}=\frac{\varepsilon_g-1+e^{-2qa}R_{g(N-1)}(2-\varepsilon_g)}{\varepsilon_g-(\varepsilon_g-1)e^{-2qa}R_{g(N-1)}}.
\label{recrel}
\end{equation}

For $1<n<N$ the boundary conditions at the surface of the layer $n$ can be written in the matrix form
\begin{equation}
\left(
\begin{array}{c}
v_{n+1}\\
w_{n+1}
\end{array}\right)=\mathbf{A}
\left(\begin{array}{c}
v_n\\
w_n
\end{array}\right).
\end{equation}

where
\begin{equation}
\mathbf{A}=\left(\begin{array}{cc}
e^{-qa}(2-\varepsilon_g) &e^{qa}(\varepsilon_g-1)\\
e^{-qa}(1-\varepsilon_g)&e^{qa}\varepsilon_g
\end{array}\right).
\end{equation}
The  eigenvectors of the matrix $\mathbf{A}$ are determined by equation
\begin{equation}
\left(\begin{array}{cc}
e^{-qa}(2-\varepsilon_g)-\lambda &e^{qa}(\varepsilon_g-1)\\
e^{-qa}(1-\varepsilon_g)&e^{qa}\varepsilon_g-\lambda
\end{array}\right)\left(\begin{array}{c}
v_n\\
w_n
\end{array}\right)=0
\end{equation}
and the eigenvalues are determined by
\begin{equation}
\left| \begin{array}{cc}
e^{-qa}(2-\varepsilon_g)-\lambda &e^{qa}(\varepsilon_g-1)\\
e^{-qa}(1-\varepsilon_g)&e^{qa}\varepsilon_g-\lambda
\end{array}\right|=\lambda^2-[2e^{-qa}+\varepsilon(e^{qa}-e^{-qa})]\lambda+1=0.
\label{eigenvalue}
\end{equation}
The roots of Eq.(\ref{eigenvalue}) are given by $\lambda_1=\lambda$  and $\lambda_2=1/\lambda$ where
\begin{equation}
\lambda=e^{-qa}+\varepsilon_g\frac{e^{qa}-e^{-qa}}{2}-\sqrt{\left[e^{-qa}+\varepsilon_g\frac{e^{qa}-e^{-qa}}{2}\right]^2-1},
\label{root}
\end{equation}
$|\lambda|<1$, $|1/\lambda|>1$. 
The eigenvectors can be written in the form
\begin{equation}
\mathbf{P_1}=\left(
\begin{array}{c}
1 \\
R_1
\end{array}\right),\,\mathbf{P_2}=\left(
\begin{array}{c}
1 \\R_2
\end{array}\right)
\end{equation}
where
\begin{equation}
R_1=\frac{ e^{-qa}-\lambda}{ e^{qa}-\lambda},\,\,R_2=\frac{ \lambda e^{-qa}-1}{ \lambda e^{qa}-1}
\end{equation}

The general solution of Eqs.(\ref{Radbouncond}) can be written in the form
\begin{equation}
\left(\begin{array}{c}
v_n\\
w_n
\end{array}\right)=C_1\mathbf{P_1}\lambda^n+C_2\mathbf{P_2}\lambda^{-n}.
\label{gensol}
\end{equation}

The boundary condition at $n=1$ can be written in the form
\begin{equation}
\left(
\begin{array}{c}
1\\
R_g\\
\end{array}\right)=\left(
\begin{array}{cc}
\varepsilon_g &(1-\varepsilon_g)\\
(\varepsilon_g-1)&(2-\varepsilon_g)
\end{array}\right)\left(\begin{array}{c}
v_1\\
w_1
\end{array}\right)=C_1\left(\begin{array}{c}
e^{-qa}\\
e^{qa}R_1
\end{array}\right)+C_2\left(\begin{array}{c}
e^{-qa}\\
e^{qa}R_2
\end{array}\right)
\label{bc1}
\end{equation}
and the boundary condition for $n=N$ has the form
\begin{equation}
C_1\lambda^{2N}R_1+C_2R_2=0.
\label{bc2}
\end{equation}
From Eqs.(\ref{bc1}) and (\ref{bc2}) we get
\begin{equation}
R_{gN}=\frac{e^{2qa}R_1R_2(1-\lambda^{2N})}{R_2-R_1\lambda^{2N}}
\label{RgN}
\end{equation}
In the limit  $N\rightarrow\infty$
\begin{equation}
R_{g\infty}=\lim_{N\rightarrow\infty}R_{gN}=e^{qa}R_1
\label{rginfty}
\end{equation}
In the limit  $qa\rightarrow\infty$ Eq.(\ref{rginfty}) is reduced to the reflection amplitude for monolayer graphene
\begin{equation}
\lim_{qa\rightarrow\infty}R_{g\infty}=R_{g1}=\frac{\varepsilon_g-1}{\varepsilon_g}.
\end{equation}

\vskip 0.5cm

The reported study was funded by RFBR according to the research project N\textsuperscript{\underline{o}} 19-02-00453

\vskip 0.5cm

$^*$alevolokitin@yandex.ru

\end{document}